# Investigation of physical dose enhancement in core-shell magnetic gold nanoparticles with TOPAS simulation

Xiaohan Xu[1,2], Yaoqin Xie[2], Jianan Wu[1,2], Zhitao Dai[1], Rui Hu[4], Luhua Wang[1,3,*]

[1] Department of Radiation Oncology, National Cancer Center/National Clinical Research Center for Cancer/Cancer Hospital & Shenzhen Hospital, Chinese Academy of Medical Sciences and Peking Union Medical College, Shenzhen, 518116, China
[2] Institute of Biomedical and Health Engineering, Shenzhen Institutes of Advanced Technology, Chinese Academy of Sciences, Shenzhen, 518055, China
[3] Department of Radiation Oncology, National Cancer Center/National Clinical Research Center for Cancer/Cancer Hospital, Chinese Academy of Medical Sciences and Peking Union Medical College, Beijing, 100021, China
[4]Department of Radiation Oncology, Affiliated Suzhou Hospital of Nanjing Medical University, Suzhou 215000, China
*Corresponding authors.

E-mail: wlhwq@yahoo.com

## Abstract

The application of metal nanoparticles as sensitization materials is a common strategy that is used to study dose enhancement in radiotherapy. Recent in vitro tests have revealed that magnetic gold nanoparticles can be used in cancer therapy under a magnetic field to enhance the synergistic efficiency in radiotherapy and photothermal therapy. However, magnetic gold nanoparticles have rarely been studied as sensitization materials. In this study, we obtained further results of the sensitization properties of magnetic gold nanoparticles using the Monte Carlo method TOPAS and TOPAS-nBio. We analyzed the properties of magnetic gold nanoparticles in monoenergetic photons and brachytherapy, and we investigated whether the magnetic field contributes to the sensitization process. Our results demonstrated that the dose enhancement factor of the magnetic gold nanoparticles was 16.7% lower than that of gold nanoparticles in a single particle irradiated by monoenergetic photons. In the cell model, the difference was less than 8.1% in the cytoplasm. We revealed that the magnetic field has no detrimental effect on radiosensitization. Moreover, the sensitization properties of magnetic gold nanoparticles in a clinical brachytherapy source have been revealed for the first time.

## 1. Introduction

Cancer is a serious disease that continues to threaten human health. At present, more than 50% of cancer patients have been treated and cured by radiotherapy [1, 2]. Although radiotherapy can kill tumor cells, it simultaneously threatens healthy tissues. Therefore, simulation studies on improving the sensitivity of tumor cells to radiotherapy and minimizing the mortality of healthy cells to enhance the efficiency of radiotherapy can provide a theoretical basis for promoting the clinical application of radiotherapy.

With the rapid developments in biotechnology and nanotechnology, the use of nanomaterials as radiosensitization materials offers new possibilities for cancer radiotherapy [3, 4, 5]. Nanoparticles (NPs) congregate in tumors as a result of enhanced permeability and retention (EPR) [6, 7]. In radiotherapy, high atomic number (Z) materials can be used to enhance the dose in tumors in combination with the EPR property. Gold Nanoparticles (AuNPs) have exhibited high X-ray cross section, low toxicity, good biocompatibility, and easy synthesis, thereby attracting significant attention in research on the radiation sensitization of nanomaterials in recent years.

AuNPs have the potential to be applied to medical imaging, medical drug delivery, photothermal therapy, and radiation sensitization therapy. In 2004, Hainfeld [8] demonstrated the radiation dose enhancement effect of AuNPs through animal experiments, which laid the foundation for research on AuNPs in radiation sensitization.

Recent advances have revealed the high potential of targeted magnetic NPs in radiotherapy [9], whereby a magnetite core combined with a suitable coating can be bestowed with biochemical and drug-delivery properties [10]. For this reason, a magnetite core combined with a gold shell was proposed to improve the stabilization, biocompatibility, and surface reactivity of sensitizing NPs [11].

The Monte Carlo (MC) method [12] is a computational approach that represents physical processes by simulating numerous random particles. Commonly used MC codes include Geant4, MCNP, and Fluka, which have a high calculation efficiency. Among these, the Geant4-DNA extension package can be used to simulate the interaction of eV energy electrons. This package has attracted the attention of medical physicists owing to its user-friendly operation interface in the form of TOPAS. TOPAS-nBio is an extension of TOPAS that is based on and extends the Geant4 Simulation Toolkit for radiobiology applications [13]. Since 2020, human lives have changed dramatically as a result of the COVID-19 pandemic. In 2021, Francis [14] investigated the influence of ionizing radiation on SARS-CoV-2 using Geant4-DNA, thereby providing a new concept for the production of an inactivated vaccine, which is still being developed at present, and revealing the extensive application prospects and significant potential of the MC method.

Nevertheless, magnetic gold nanoparticles have rarely been studied as sensitization materials. To address this limitation, in this study, we used TOPAS [15, 16] and TOPAS-nBio [17] to study the $Fe_3O_4$@AuNP properties in radiotherapy sensitization compared to an AuNP in a single nanoparticle and a cell model using monoenergetic photons. Subsequently, we combined the simulation with a magnetic field to investigate the influence on the sensitivity process. Finally, we changed the photon beams with a brachytherapy source to perform the same process. Our work contributes to the research on $Fe_3O_4$@AuNPs in radiotherapy using the MC method and provides a reference for clinical research.

## 2. Materials and methods

### 2.1 Cell uptake of $Fe_3O_4$@AuNPs by HeLa cells with or without magnetic field

The $Fe_3O_4$@AuNP that was used in the test of Hu [18] consisted of a $Fe_3O_4$ core and a gold shell, as shown in Fig. 1(a). The mean diameter of the $Fe_3O_4$@AuNPs was 100 nm according to dynamic light scattering analysis, as illustrated in Fig. 1(b). According to Figs. (a) and (b), we obtained that the $Fe_3O_4$@AuNP consisted of a 60 nm diameter $Fe_3O_4$ core and a 20 nm thickness gold shell. Therefore, we selected 100 nm as the diameter of the $Fe_3O_4$@AuNP and used the same $Fe_3O_4$ and Au ratio in our simulation work. Hu used confocal laser scanning microscopy (CLSM) to observe the distribution of the $Fe_3O_4$@AuNPs internalized by the HeLa cells, as depicted in Fig. 1(c). The $Fe_3O_4$@AuNPs and lysosomes were labeled by fluorescein isothiocyanate and LysoTracker Red, and exhibited green and red fluorescence, respectively, in the CLSM. The distributions of the $Fe_3O_4$@AuNPs and lysosomes were clearly partially overlapped. It meant the $Fe_3O_4$@AuNPs were internalized by cells and could be swallowed by the lysosomes in the cytoplasm. We also observed that the green fluorescence intensity with a 0.2 T magnetic field was higher than that without a magnetic field. Hu used flow cytometry to analyze the mean fluorescence intensity to compare the cell uptake of $Fe_3O_4$@AuNPs with and without a magnetic field quantitatively, as illustrated in Fig. 1(d). The results demonstrated that the fluorescence intensity of the $Fe_3O_4$@AuNPs in an external magnetic field was 1.48 times higher than that without a magnetic field. Moreover, Hu demonstrated that $Fe_3O_4$@AuNPs can be used to decrease the viability of HeLa cells in radiotherapy with an external magnetic field (0.2 T). The results indicated that the cell viability was affected by the magnetic field owing to the cell uptake properties being enhanced under the magnetic field. In fact, the cell viability may be affected by the cell uptake properties, the physical dose enhancement of $Fe_3O_4$@AuNPs, and other conditions. In our research, we studied the dose enhancement properties of $Fe_3O_4$@AuNPs with and without a magnetic field using TOPAS and TOPAS-nBio.

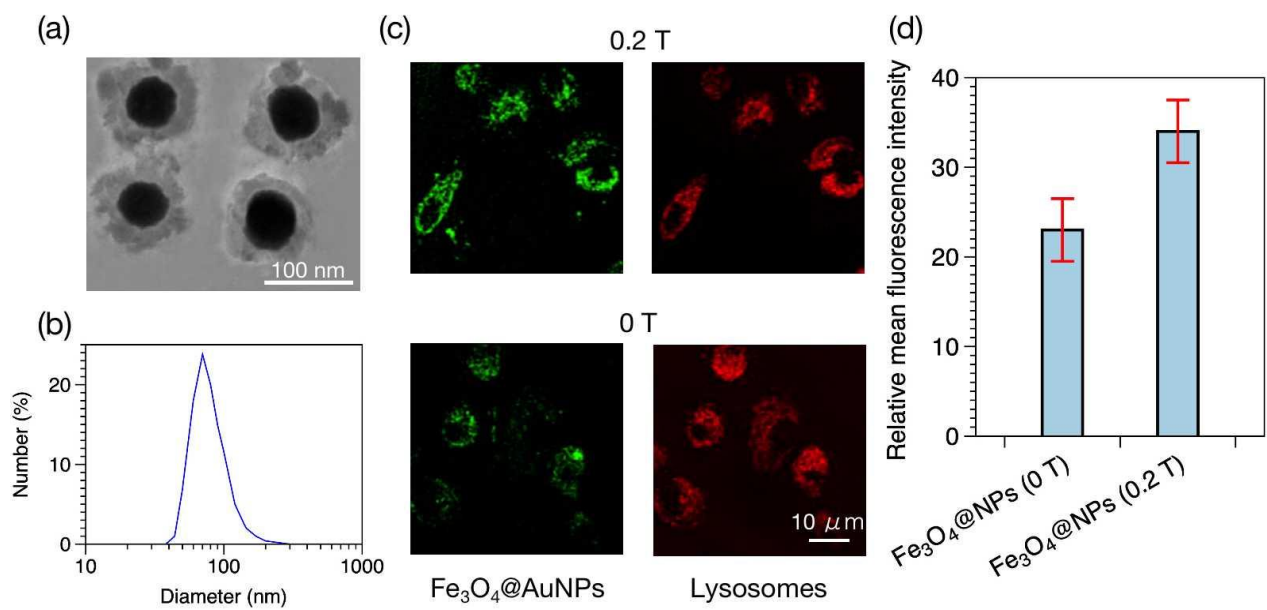

Figure 1. (a) TEM image of $Fe_3O_4$@AuNPs. (b) Dynamic light scattering analysis of $Fe_3O_4$@AuNPs. (c) Distribution of $Fe_3O_4$@AuNPs in HeLa cells with and without magnetic field in CLSM after 3 h of incubation. (d) Fluorescence intensity of $Fe_3O_4$@AuNPs with and without magnetic field.

## 2.2 Calculation methods for dose enhancement factor: two-step and one-step methods

Two methods are used for calculating the dose enhancement factor (DEF) with TOPAS and TOPAS-nBio. Lin [19] investigated the dose enhancement of proton and photon irradiation on AuNPs using TOPAS. Lin calculated the dose distribution around a single AuNP that changed with the distance from the particle surface and obtained the distribution of the DEF at different distances from the AuNP surface. The DEF is defined as the ratio of the dose deposited around the metal nanoparticle (MNP) and water nanoparticle (WNP). As the Geant4-DNA physics processes are workable in water only, this package cannot be used to calculate the tracks in AuNP accurately; thus, Lin divided the dose calculation into two steps. First, the physics module Penelope was activated to calculate the interaction between the particle source and AuNP in water, following which the generated secondary electrons that were excited from the AuNP surface were stored in the phase space file. Second, the phase space file was placed into a water box to calculate the dose distribution of the secondary electrons in water and Geant4-DNA was activated in the water box region. Such a method of calculating the DEF is referred to as the "two-step method" in our research. However, the surface dose distribution around a single AuNP is not the exclusive factor affecting the cell livability, and the effects of the radiation emerging or scattering from an AuNP on the other AuNPs in a cell model should also be considered.

Scientists developed TOPAS-nBio to simulate radiobiological experiments on nanometer scale cells considering the physics, chemistry, and biology effects. TOPAS-nBio supports the assignment of different physical models to different geometry components. Rudek [20] established the AuNPs that were internalized in a cell model irradiated by photons, protons, and carbon ions respectively using TOPAS-nBio. To define suitable physical modules in different regions, Rudek set the Livermore physical process in the AuNP region and the Geant4-DNA process outside the AuNP region. Thereafter, the DEFs in the cytoplasm and nucleus were calculated. This method of calculating the DEF is referred to as the "one-step method" in this work.

The two methods mentioned above are aimed at a single nanoparticle and a single cell. A single cell includes the physical interaction between the primary beam and MNPs. Therefore, in the simulation study of $Fe_3O_4$@AuNPs, we considered the calculation results of both the two-step and one-step method to evaluate the sensitivity enhancement performance in a single nanoparticle as well as in a cell. To compare the two-step method and one-step method, we modeled the same geometry to simulate the interaction process of the photons and AuNP in water, as illustrated in Fig. 2 (a).

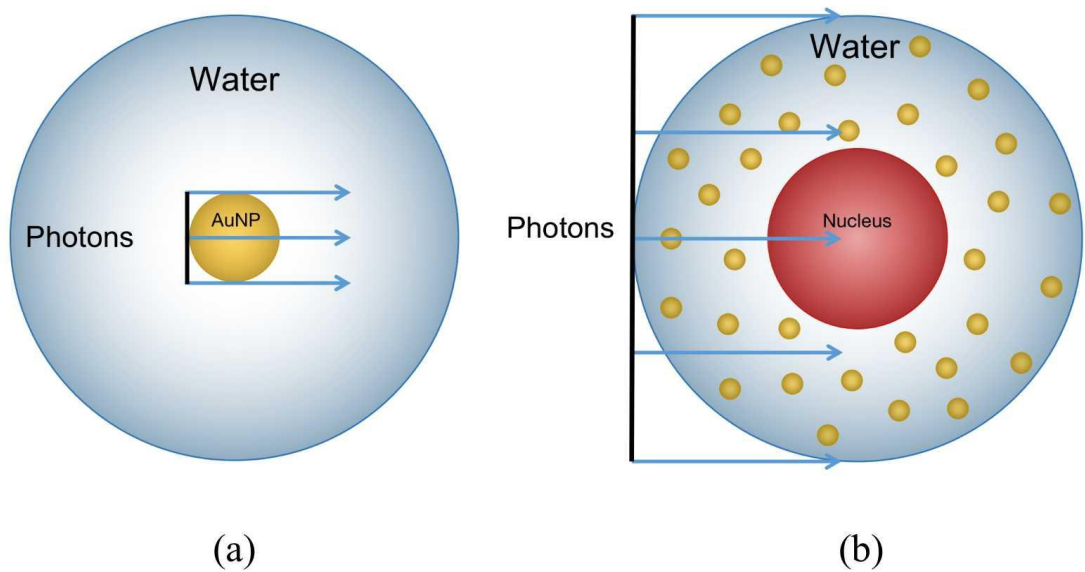

Figure 2. Geometry sketch of MC simulation (not the actual scale): (a) A 100 nm diameter AuNP was placed in a 20 μm diameter sphere filled with water. A photon beam with the same size as the AuNP in diameter was placed upstream to the AuNP. The dose was scored in the grids with a volume size of $1 \times 1 \times 1$ nm$^3$ for 0 to 150 nm, $10 \times 10 \times 10$ nm$^3$ for 150 nm to 1.95 μm, and $100 \times 100 \times 100$ nm$^3$ for 1.95 μm to 19.95 μm from the AuNP surface. (b) The 10 μm diameter water sphere contained a 5 μm water sphere in the center to model the cell containing a nucleus. The 100 nm diameter NPs were randomly distributed in the cytoplasm. The photon beam had the same diameter as the cell and was placed upstream to the cell.

For the two-step method, we divided the simulation into two steps, as described above. In the first step, the AuNP was placed in a box ($100 \times 100 \times 100$ nm$^3$) filled with vacuum and irradiated by a 50 keV photon beam, following which the output electron phase space file was obtained from the AuNP surface. In the second step, the electron phase space file was used as a particle source and placed in the center of the water sphere (20 μm diameter). The deposited dose was scored in the grids from 0 to 150 nm, 150 nm to 1.95 μm, and 1.95 μm to 19.95 μm from the AuNP surface with different precisions. We set the Livermore physics processes for the first step and the Geant4-DNA physics for the second step. The de-excitation was activated to include Auger production and particle induced X-ray emission.

In the one-step method, the AuNP was placed in the center of the water box (20 μm diameter). Thereafter, the 50 keV photon beam interacted with the AuNP and the dose was recorded at different distances from the AuNP surface. The AuNP region was assigned with Livermore physics processes, whereas all of the other regions were set with Geant4-DNA physics processes. In both methods, we recorded the dose distribution that was produced by the electrons and photons to investigate which particle types mainly contributed to the deposited dose.

### 2.3 Photon energy dependence of single Fe$_3$O$_4$@AuNP dose enhancement using two-step method

We used an Fe$_3$O$_4$@AuNP with the same size and composition as in Hu's test in our simulation. Five monoenergetic photon beams (50, 100, 150, 200, and 250 keV) were used as particle sources to irradiate the single Fe$_3$O$_4$@AuNP, AuNP, and WNP. The photon source was plane parallel with a 100 nm diameter and started at the nanoparticle surface, as illustrated in Fig. 2(a). To evaluate the properties of the Fe$_3$O$_4$@AuNP at different photon energies, we compared the DEFs of the Fe$_3$O$_4$@AuNP and AuNP that were irradiated by the same five monoenergetic photon beams with the same simulation parameters.

### 2.4 Photon energy and Nanoparticles concentration dependence of cell dose enhancement using one-step method

The radiation processes were implemented in a simplified cell model. The 10 μm diameter cell contained a 5 μm diameter nucleus in the center and the cell was placed in a water box. Both the cytoplasm and nucleus were filled with water to model the cellular environment. The monoenergetic photon source (50, 100, 150, 200, and 250 keV) was plane parallel with a 10 μm diameter and started from the cell surface, as illustrated in Fig. 2(b). Considering that NPs are predominantly dispersed in the cytoplasm when NPs enter the cell [21], the 100 nm diameter $Fe_3O_4$@AuNPs and 100 nm diameter AuNPs were placed in the cytoplasm randomly respectively in the simulation to draw a comparison. To cover the desired dose range on the cell level, the NPs mass concentration was incremented in the range of 1 to 50 mg/mL [22]. Subsequently, we selected 1, 5, 10, and 50 mg/mL as the concentration weights of the $Fe_3O_4$@AuNPs and AuNPs in the cytoplasm. The corresponding NPs numbers are listed in Table 1.

Table 1. Number of 10 μm diameter $Fe_3O_4$@AuNPs and AuNPs in cytoplasm for five concentration weights (units: mg/mL).

| Mass/volume (mg/mL) | 1 | 5 | 10 | 50 |
|---|---|---|---|---|
| Number of $Fe_3O_4$@AuNPs in cytoplasm | 62 | 307 | 615 | 3074 |
| Number of AuNPs in cytoplasm | 52 | 259 | 518 | 2588 |

## 2.5 Magnetic field dependence of single nanoparticle and cell dose enhancement

With the increasing use of MRI-guided radiotherapy, it is necessary to investigate the influence of the magnetic field on the radiotherapy. The in vitro tests performed by Hu [18] demonstrated that core-shell $Fe_3O_4$@AuNPs can be used to decrease the viability of HeLa cells by improving their internalization by the cells in an external magnetic field (0.2 T). Bug [23] and Lazarakis [24] demonstrated that the magnetic field affected the charged particle trajectory only; the physical cross section, DNA strand breaks, and cluster size distribution could not be changed by the magnetic field in Geant4.

Scientists have shown that magnetic targeting is a promising technology among passive tumor accumulation in radiotherapy. Magnetic NPs can be focused on the tumors under the magnetic field outside the body [25]. However, the magnetic targeting property for magnetic material in a magnetic field cannot be simulated with the MC method. Therefore, we used four concentrations of $Fe_3O_4$@AuNPs to simulate the targeting focus of the $Fe_3O_4$@AuNPs in four magnetic fields, as discussed in section 2.4.

We investigated the influence of the changed particle trajectory under the magnetic field on the sensitization process of the $Fe_3O_4$@AuNP and AuNP. The simulation was performed on a single nanoparticle and a cell model using the two-step and one-step methods, with irradiation by a 50 keV monoenergetic photon beam. The models used were the same as those described in sections 2.3 and 2.4. An external magnetic field with a strength of 0.1 to 2 T was used in the simulations. The NPs concentration was 50 mg/mL in the cell model.

## 2.6 DEFs of $Fe_3O_4$@AuNP and AuNP interacted with brachytherapy source

In the in vitro tests of Hu [18], the HeLa cells were irradiated by photons from a Varian linear accelerator (True Beam). In this study, we further evaluated the sensitization properties of the $Fe_3O_4$@AuNP and AuNP under a clinically applied source. We implemented the Varian GammaMed Plus HDR $^{192}$Ir brachytherapy source model [26] using TOPAS to explore the DEF of the $Fe_3O_4$@AuNP irradiated by a brachytherapy source. The particle numbers that were emitted from the brachytherapy source model per keV for the per initial photons, which were recorded on a parallel plane at a 2 cm distance from the source center, are presented in Fig. 3. A total of $10^8$ initial photons were used as the source beam and the source was placed in an 80 × 80 × 80 cm$^3$ water box to calculate the dose distribution in water. With the exception of the radiation source, all parameters were consistent with those described in section 2.5.

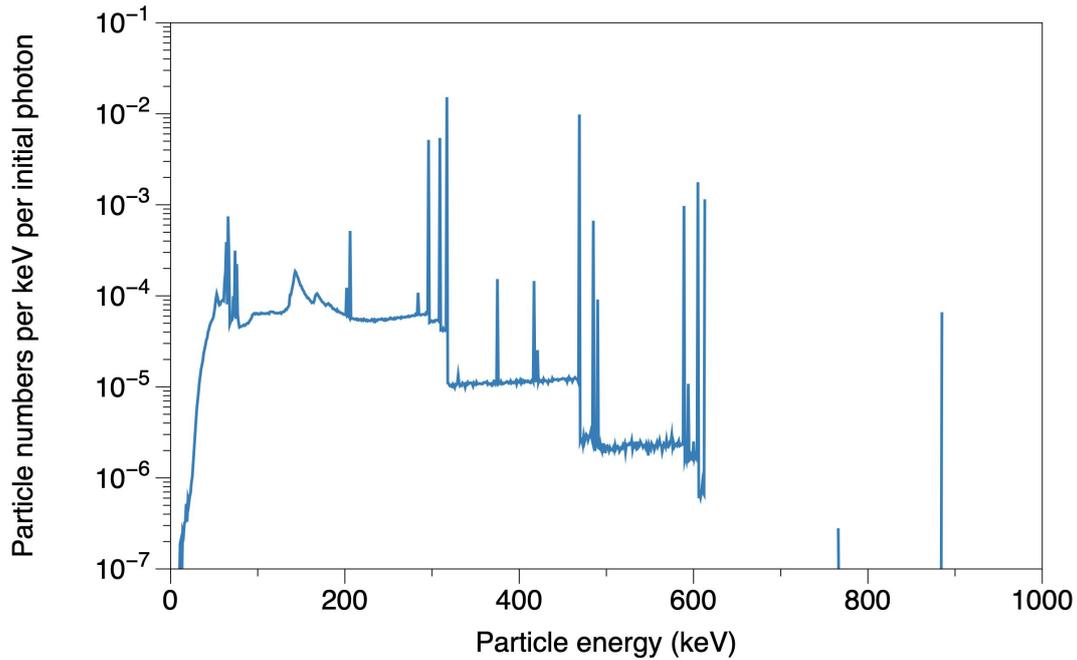

Figure 3. Particle energy spectra emitted from brachytherapy source model, recorded on parallel plane at 2 cm distance from source center.

## 3. Results

### 3.1 Comparison of two-step and one-step methods

Figure 4 presents the results of the comparison between the one-step and two-step methods as well as the dose contributions from the electrons and photons. It is obvious that the four curves in Fig. 4 exhibit similar trends. The results reveal two significant conclusions. First, the deposited dose was mainly contributed by the secondary electrons. The dose produced by the photons was very slight compared to the electrons in both the one-step and two-step methods. According to the results, only the electrons need to be considered in irradiation simulation to improve the calculation efficiency, regardless of whether the one-step or two-step method is used. Second, the dose distributions of the one-step and two-step methods exhibited no significant differences. Thus, we can use the two-step method to calculate the DEF around a single nanoparticle and the one-step method to calculate the DEF in a cell model. Moreover, the influence of the two methods on the simulation results does not need to be considered.

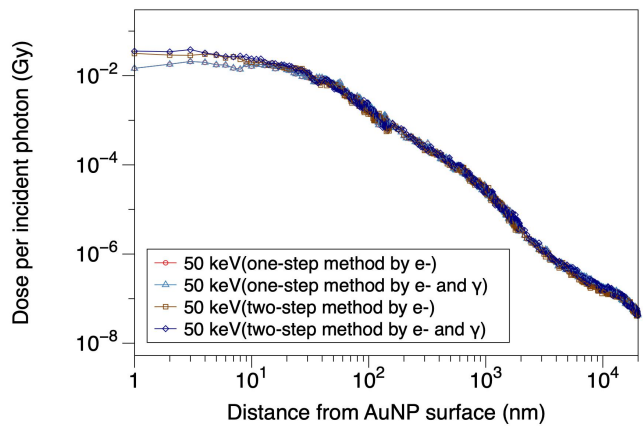

Figure 4. Dose distribution per incident photon vs. distance from AuNP surface.

### 3.2 Photon energy dependence of single Fe$_3$O$_4$@AuNP dose enhancement

The results of the photon irradiations are depicted in Fig. 5. Figures 5(a), (b), and (c) present the dose distributions at different distances from the surface of the single Fe$_3$O$_4$@AuNP, single AuNP, and single WNP, respectively, per incident photon. It is clear that the five dose distribution curves in both Figs. 5(a) and (b) exhibited similar trends. Higher energy of the photon led to a higher dose distribution in the energy range from 150 to 250 keV. However, the deposited dose of the 100 keV photon was higher than 50 keV within the range of $1.3 \times 10^3$ to $7 \times 10^3$ nm. According to Fig. 5(c), the 100 keV photon dose distribution was higher than that of the 50 keV photon in the range of $1 \times 10^3$ to 1.4

× $10^4$ nm. The DEFs of the $Fe_3O_4$@AuNP and AuNP were calculated based on Figs. 5(a), (b), and (c) and the results are plotted in Figs. 5(d) and (e). The five curves in Figs. 5(d) and (e) also exhibited similar trends. The DEFs of both NPs increased with an increase in the photon energy for the 150, 200, and 250 keV photons. The curves of the 50 and 100 keV photons crossed at $10^3$ and $1.8 \times 10^4$ nm. To compare the total dose deposition in the range of 1 to $2 \times 10^4$ nm intuitively, the doses that were distributed at different distances were totaled for each photon energy, as illustrated in Fig. 5(f). According to the figure, the DEF of the $Fe_3O_4$@AuNP was 16.7% lower than that of the AuNP on average. Moreover, the peak of the DEF verus photon energy curve appeared at 100 keV. For thousands of keV energy photons irradiated with $Fe_3O_4$@AuNP and AuNP, the DEF was the highest near 100 keV.

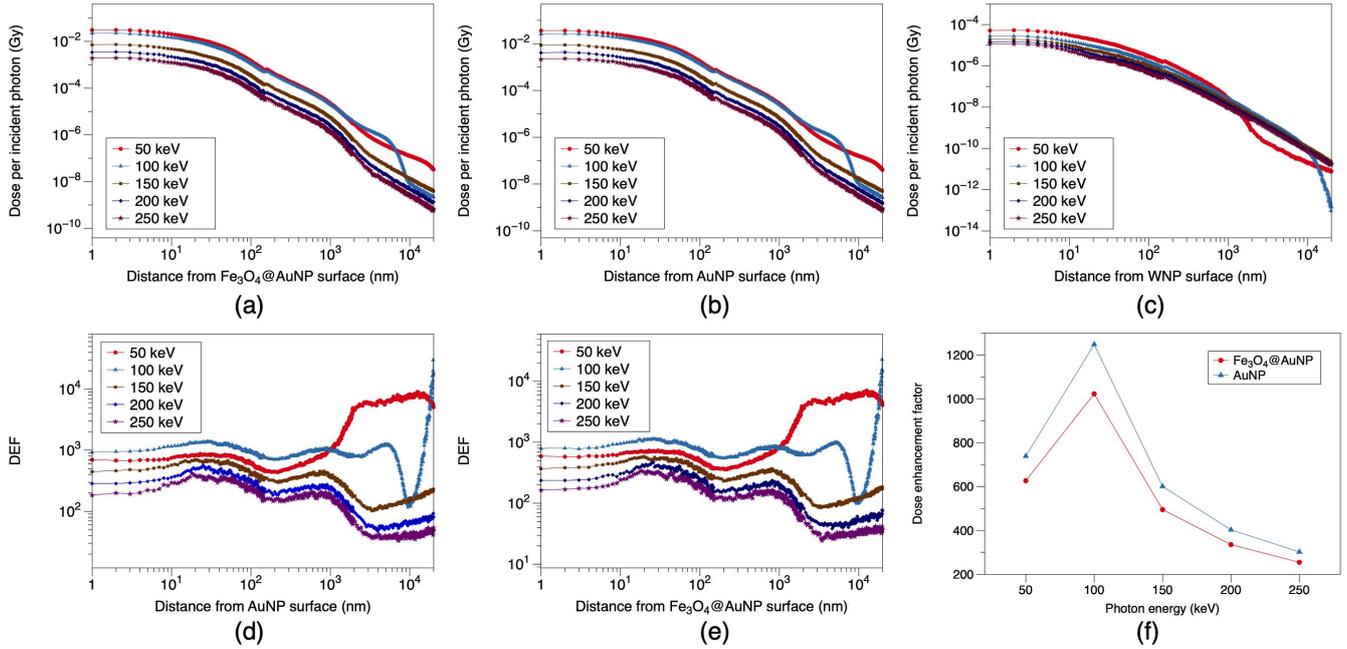

Figure 5. Relationship between dose distribution per incident photon and distance from surface of (a) $Fe_3O_4$@AuNP, (b) AuNP, and (c) WNP for 50, 100, 150, 200, and 250 keV photons. DEF distributions around (d) $Fe_3O_4$@AuNP and (e) AuNP as function of distance from the nanoparticle surface. (f) Total DEF around $Fe_3O_4$@AuNP and AuNP in range of 1 to $2 \times 10^4$ nm vs. photon energy.

## 3.3 Photon energy and Nanoparticles concentration dependence of cell dose enhancement

The DEFs of the $Fe_3O_4$@AuNPs and AuNPs in the cytoplasm and nucleus are presented in Figs. 6(a) and (b), respectively. In the cytoplasm, the DEFs of the $Fe_3O_4$@AuNPs and AuNPs decreased with an increase in the photon energy. The maximum DEFs of the $Fe_3O_4$@AuNPs and AuNPs for 50 mg/mL were 3.69 and 3.83. The maximum difference was within 2.1%, 1%, 2.2%, and 8.1% when comparing the DEFs of the AuNPs and the $Fe_3O_4$@AuNPs for the 1, 5, 10, and 50 mg/mL NPs concentrations, respectively. In the nucleus, the DEF of the AuNPs reached the maximum at 100 keV with the 5, 10, and 50 mg/mL weight concentrations and 50 keV with 1 mg/mL. The maximum difference was within 1.9%, 5%, 13%, and 3.1% when comparing the AuNPs and $Fe_3O_4$@AuNPs for the 1, 5, 10, and 50 mg/mL NP concentrations, respectively.

The DEF differences between the $Fe_3O_4$@AuNPs and AuNPs in the cytoplasm and nucleus are summarized in Table 2. It can be observed that the energy had a greater influence on the DEF when the particle concentration was higher. In the cytoplasm, the DEF of the AuNPs was higher than that of the $Fe_3O_4$@AuNPs except for 1 mg/mL. However, the difference was not obvious in the nucleus. Furthermore, in general, a higher NPs concentration led to a higher DEF in the cytoplasm and nucleus. This means that the high magnetic focus property can achieve better dose enhancement for radiotherapy.

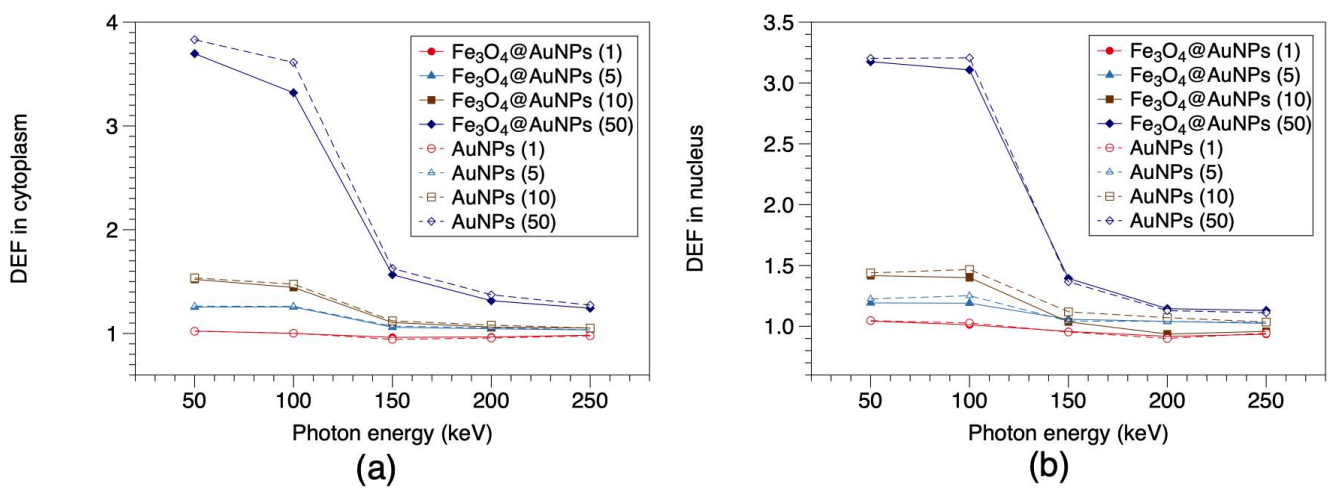

Figure 6. Function of DEF and photon energy in (a) cytoplasm and (b) nucleus.

Table 2. DEF differences between Fe$_3$O$_4$@AuNPs and AuNPs in cytoplasm and nucleus with different photon energies and NPs concentrations.

| Photon energy (keV) | Concentration in weight (mg/mL) | | | | | | | |
|---|---|---|---|---|---|---|---|---|
| | 1 | 5 | 10 | 50 | 1 | 5 | 10 | 50 |
| | DEF differences between AuNPs and Fe$_3$O$_4$@AuNPs in cytoplasm (%) | | | | DEF differences between AuNPs and Fe$_3$O$_4$@AuNPs in nucleus (%) | | | |
| 50 | -0.16 | 0.68 | 0.90 | 3.51 | 0.34 | 2.67 | 1.53 | 0.87 |
| 100 | 0.19 | 0.41 | 2.11 | 8.09 | 1.45 | 4.91 | 4.56 | 3.09 |
| 150 | -2.05 | 0.98 | 1.46 | 3.74 | -0.50 | -1.40 | 7.35 | -2.04 |
| 200 | -1.09 | 1.00 | 1.76 | 4.19 | -1.88 | 0 | 12.54 | -1.45 |
| 250 | -0.65 | 0.15 | -0.03 | 2.36 | 0.98 | 0.67 | 7.43 | -1.71 |

## 3.4 Magnetic field dependence of dose enhancement

The relationship between the magnetic field and DEF of the NPs is presented in Figs. 7 and 8. We used the two-step method on a single nanoparticle and found that the DEF of the AuNP was 15% higher than that of the Fe$_3$O$_4$@AuNP under the magnetic field, as illustrated in Fig. 7. The DEF was stable above 0.5 T and the value was slightly higher than the DEF without a magnetic field. The minimum DEF value appeared at 0.2 T and this was 2.5% lower than the constant value. This simulation result indicates that the magnetic field did not contribute significantly to the DEF.

The DEFs of the AuNP and Fe$_3$O$_4$@AuNP in the cytoplasm and nucleus are illustrated in Fig. 8. We used the one-step method in a cell to simulate the influence of the magnetic field on the DEF. The Fe$_3$O$_4$@AuNP DEF was 3.9% and 3.1% lower than that of the AuNP in the cytoplasm and nucleus, respectively. The DEF in the cytoplasm was 11.7% and 12.4% higher than that in the nucleus for the Fe$_3$O$_4$@AuNP and AuNP. In general, the magnetic field did not contribute significantly to the DEF in the cell model. In this study, we concluded that a magnetic field with a strength of 0.1 T to 2 T would not have a negative effect on the sensitization process.

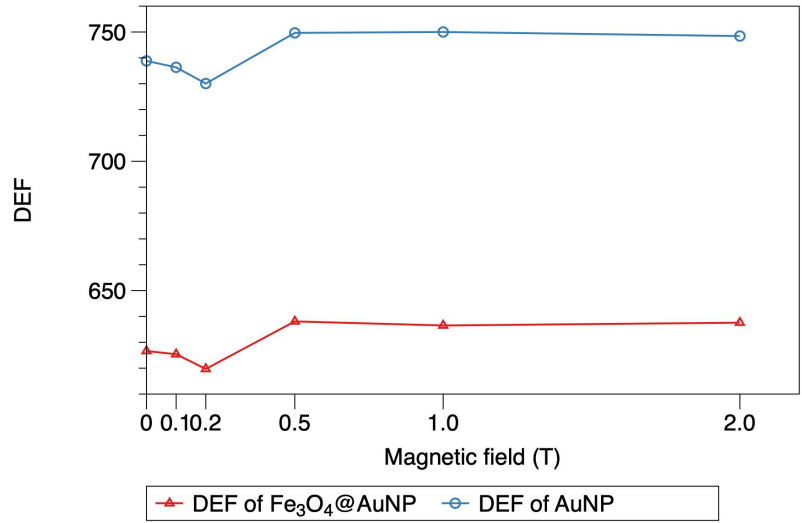

Figure 7. Relationship between magnetic field and DEF of a single Fe$_3$O$_4$@AuNP and AuNP.

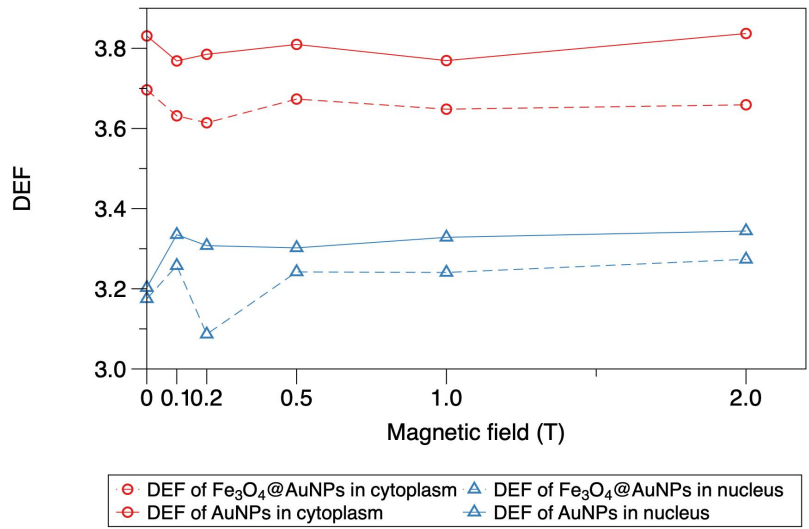

Figure 8. Relationship between magnetic field and DEF of Fe$_3$O$_4$@AuNPs and AuNPs in cytoplasm and nucleus.

## 3.5 DEFs of Fe$_3$O$_4$@AuNP and AuNP irradiated by brachytherapy source

The results of the brachytherapy source irradiations are depicted in Fig. 9. For the single nanoparticle model in Fig. 9(a), the DEF of the Fe$_3$O$_4$@AuNP was 9.26% lower than that of the AuNP. For the cell model in Fig. 9(b), the DEFs of the Fe$_3$O$_4$@AuNP were 6.3% and 2.7% lower than those of the AuNP in the cytoplasm and nucleus, respectively, whereas the DEFs of the Fe$_3$O$_4$@AuNP and AuNP in the cytoplasm were 26.75% and 31.62% higher than those in the nucleus.

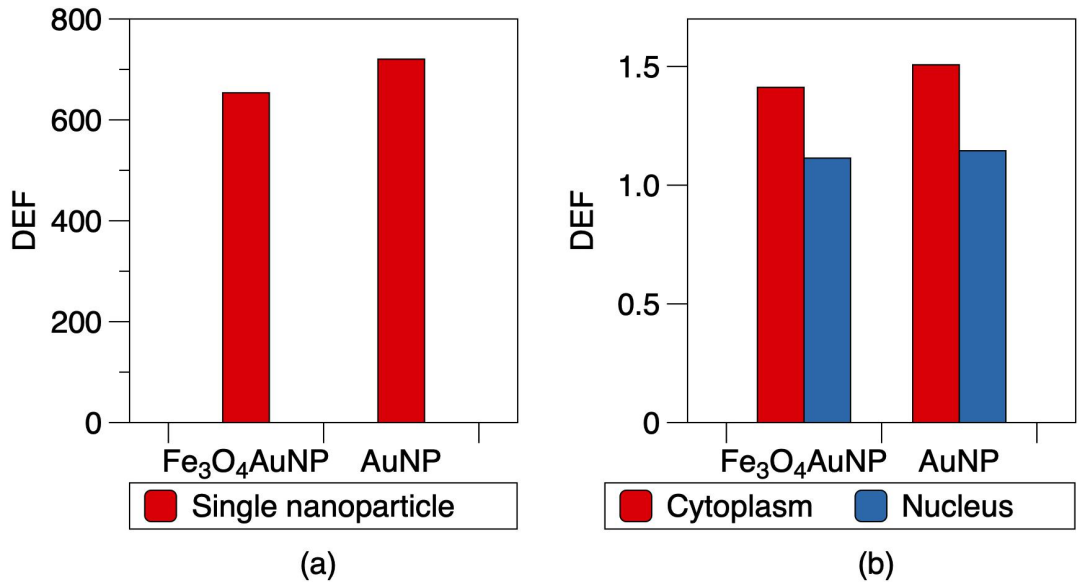

Figure 9. Total DEF of Fe$_3$O$_4$@AuNP and AuNP in (a) single nanoparticle in range of 1 to $2 \times 10^4$ nm, and (b) in cytoplasm and nucleus when irradiated by brachytherapy source.

## 4. Discussion

AuNPs are studied extensively in radiosensitization owing to their properties of high X-ray absorption, hypotoxicity, and easy synthesis. Magnetite can be used as targeting material to improve tumor drug delivery because of the magnetic targeting property in the magnetic field [25]. As a novel nanoparticle, the Fe$_3$O$_4$@AuNP combines the properties of gold and magnetite, and it has been used in in vitro experiments to decrease the cell survival rate [18].

In this study, we explored the DEF of an Fe$_3$O$_4$@AuNP in a single nanoparticle and in a cell model compared to the AuNP. The DEF around the single Fe$_3$O$_4$@AuNP was 16.7% lower than that of the AuNP, and the differences between the AuNP and Fe$_3$O$_4$@AuNP in the cytoplasm and nucleus are detailed in Table 2. With the increase in the NPs concentration, the DEF residuals between the Fe$_3$O$_4$@AuNPs and AuNPs increased in the cytoplasm. It was expected that the DEF of the Fe$_3$O$_4$@AuNP would be lower than that of the AuNP because the photoelectric cross section of iron and oxygen is lower than that of gold. We quantified the discrepancy between the Fe$_3$O$_4$@AuNP and AuNP to provide an analysis of core-shell magnetic NPs that are used as sensitivity materials. It is well known that the clustering property of AuNPs will decrease the DEF in a cell in radiotherapy. However, little research has been conducted on how to avoid magnetic NP clustering. Significant analytical potential exists for improving the magnetic NPs stability by modifying the extra magnetic field so as to increase the DEF.

We investigated the influence of the magnetic field on the DEF and demonstrated that the magnetic field did not have a significant effect on the sensitization process. The results revealed that the changed electron trajectory was insufficient to influence the dose enhancement, or the electron trajectory was insufficient to be changed with such electron energy and the magnetic field [23]. Therefore, the magnetic field would not risk physical enhancement because the electron energy was too high according to the magnetic field. Combined with the in vitro experiment carried out by Hu, we verified that the radiosensitization mainly benefited from the physical enhancement of Fe$_3$O$_4$@AuNP in addition to the cell uptake in the magnetic field.

Furthermore, we constructed a brachytherapy source for irradiation with a single nanoparticle and a cell model. The results of the brachytherapy irradiation showed the residuals between the Fe$_3$O$_4$@AuNP and AuNP in a single nanoparticle and a cell model. The DEF differences between the AuNP and Fe$_3$O$_4$@AuNP were 9.26%, 6.3%, and 2.7% for the single particle, cytoplasm, and nucleus, respectively. The results clarified the dose enhancement of the Fe$_3$O$_4$@AuNPs under the brachytherapy source. In the future, research on guiding the Fe$_3$O$_4$@AuNPs to focus on tumors through the magnetic field will be quite beneficial. For example, the source applicator may be magnetized to guide magnetic NPs or the sensitization may be combined with MRI-guided brachytherapy to focus the magnetic NPs. This research may raise concerns regarding MRI-guided brachytherapy combined with magnetic NPs.

## 5. Conclusions

In this work, we compared the one-step and two-step methods for calculating the DEF to verify that there was no significant difference between the methods. Thereafter, we applied the two methods to a single particle and a cell model to investigate the DEFs of the $Fe_3O_4@AuNP$ and AuNP. The DEF of the $Fe_3O_4@AuNP$ was 16.7% lower than that of the AuNP in a single particle. In the cell model, the DEF difference between the $Fe_3O_4@AuNP$ and AuNP was below 8.1% in the cytoplasm with an NPs concentration of 1 to 50 mg/mL. We also demonstrated that the magnetic field has no detrimental effect on the NPs radiosensitization. Furthermore, we applied a brachytherapy source for interaction with the $Fe_3O_4@AuNP$ and AuNP in a single nanoparticle and a cell model to obtain the DEF in brachytherapy source irradiation.

In summary, this study has revealed the $Fe_3O_4@AuNP$ properties in radiotherapy dose enhancement using the MC method for the first time. Moreover, we demonstrated that the physical dose enhancement of the $Fe_3O_4@AuNP$ is independent of the magnetic field. Finally, we determined the DEF of $Fe_3O_4@AuNPs$ in a brachytherapy source to provide simulation results for clinical research. In future research, $Fe_3O_4@AuNPs$ may be combined with a magnetic field (such as MRI) to overcome the challenge of NPs clustering and to improve the NPs concentration in the cell. This will be desirable for future in vitro tests on radiosensitization as well as clinical research.

## Acknowledgements

This work is supported by Sanming Project of Medicine in Shenzhen(No. SZSM201612063). The authors would like to thank HD Video R & D Platform for Intelligent Analysis and Processing in Guangdong Engineering Technology Research Center of Colleges and Universities(GCZX-A1409), ShenZhen University for providing access to their high performance workstations.